\documentclass{llncs}
\usepackage[T1]{fontenc} 
\usepackage{graphicx}
\usepackage[boxed,vlined,ruled,linesnumbered]{algorithm2e}
\usepackage{cite}
\usepackage{amsmath}
\usepackage{amsfonts}
 
\usepackage{color}
\usepackage{multicol}
\usepackage{multirow}
\usepackage{colortbl}
\usepackage{enumerate}
\usepackage{framed}
\usepackage[singlelinecheck=off, labelfont=bf]{caption}
\usepackage{listings}

\newcommand{\R}{\mathbb{R}}

\begin{document}

\title{Applying machine learning to the problem of choosing a heuristic to select the variable ordering for cylindrical algebraic decomposition}

\institute{
$^1$ University of Cambridge Computer Laboratory, Cambridge CB3 0FD, U.K. \\ 
$^2$ University of Bath, Department of Computer Science, Bath, BA2 7AY, U.K. \\ 
	\email{ 
		{\tt \{zh242, lp15, jpb65\}@cam.ac.uk}, \\
		{\tt \{J.H.Davenport, M.England, D.J.Wilson\}@bath.ac.uk}
	}
}

\author{
Zongyan Huang\inst{1},
Matthew England\inst{2},
David Wilson\inst{2}, \\
James H. Davenport\inst{2}, 
Lawrence C. Paulson\inst{1} 
and James Bridge\inst{1}
}

\maketitle

\begin{abstract}

Cylindrical algebraic decomposition(CAD) is a key tool in computational algebraic geometry, particularly for quantifier elimination over real-closed fields.  When using CAD, there is often a choice for the ordering placed on the variables.  This can be important, with some problems infeasible with one variable ordering but easy with another. Machine learning is the process of fitting a computer model to a complex function based on properties learned from measured data. In this paper we use machine learning (specifically a support vector machine) to select between heuristics for choosing a variable ordering, outperforming each of the separate heuristics.


\end{abstract}


\section{Introduction} 
\label{SEC:Intro}

Cylindrical algebraic decomposition (CAD) is a key tool in real algebraic geometry.  It was first introduced by Collins \cite{Collins1975} to implement quantifier elimination over the reals, but has since been applied to applications including robot motion planning \cite{WDEB13}, programming with complex valued functions \cite{DBEW12}, optimisation \cite{FPM05} and epidemic modelling \cite{BENW06}.  Decision methods for real closed fields are of great use in theorem proving \cite{DSW98b}. \textsc{MetiTarski} \cite{AP10}, for example, decides the truth of statements about special functions using CAD and rational function bounds.

When using CAD, we often have a choice over which variable ordering to use.  It is well known that this choice is very important and can dramatically affect the feasibility of a problem.  In fact, Brown and Davenport \cite{BD07} presented a class of problems in which one variable ordering gave output of double exponential complexity in the number of variables and another output of a constant size.  Heuristics have been developed to help with this choice, with Dolzmann~et~al.~\cite{DSS04} giving the best known study. However, in CICM last year \cite{BDEW13}, it was shown that even the best known heuristic could be misled. Although that paper provided an alternative heuristic, this had its own shortcomings, and it now seems likely that no one heuristic is suitable for all problems.  

Our thesis is that the best heuristic to use is dependent upon the problem considered.  However, the relationship between the problems and heuristics is far from obvious 
and so we investigate whether machine learning can help with these choices. 
Machine learning is a branch of artificial intelligence. It uses statistical methods to infer information from supplied data which is then used to make predictions for previously unseen data \cite{alpaydin2004introduction}. 
We have applied machine learning (specifically a support vector machine) to the problem of selecting a variable ordering for both CAD itself and quantifier elimination by CAD, using the nlsat dataset \cite{nlsat} of fully existentially quantified problems.  Our results show that the choices made by machine learning are on average superior to both any individual heuristic and to picking a heuristic at random.  The results also provide some new insight on the heuristics themselves.
This appears to be the first application of  machine learning to problem formulation for computer algebra, although it follows recent application to theorem proving \cite{HP13, BHP14}.

We conclude the introduction with background theory on CAD and machine learning.  Then in Sections \ref{SEC:Methodology}, \ref{SEC:Results} and \ref{SEC:Future} we describe our experiment, its results and how they may be extended in the future.
Finally in Section \ref{SEC:Conclusion} we give our conclusions and ideas for future work.

\subsection{Quantifier elimination and CAD}
\label{SUBSEC:CAD}

Let $Q_i \in \{\exists,\forall\}$ be quantifiers and $\phi$ be some quantifier free formula.  Then given
\[
\Phi(x_1, \ldots, x_k) := Q_{k+1}x_{k+1} \ldots Q_n x_n \, \phi(x_1,\ldots,x_n),
\]
\emph{quantifier elimination} (QE) is the problem of producing a quantifier free formulae $\psi(x_1,\ldots,x_k)$ equivalent to $\Phi$.  In the case $k=0$ this reduces to the \emph{decision problem}, is $\Phi$ true?  Tarski proved that QE was possible for semi-algebraic formulae (polynomials and inequalities) over $\R$ \cite{Tarski98}.  However, the complexity of Tarski's method is non-elementary (indescribable as a finite tower of exponentials) and so CAD was a major breakthrough when introduced, despite complexity doubly exponential in the number of variables.  
For some problems QE is possible through algorithms with better complexity (see for example the survey by Basu \cite{Basu11}), but CAD implementations remain the best general purpose approach.  

Collins' algorithm \cite{ACM84I} works in two stages.  First, \emph{projection} calculates sets of projection polynomials $S_i$ in variables $(x_1, \dots, x_i)$.  This is achieved by repeatedly applying a projection operator onto a set of polynomials, producing a set with one variable fewer.  We start with the polynomials from $\phi$ and eliminate variables this way until we have the set of univariate polynomials $S_1$.

Then in the \emph{lifting} stage, decompositions of real space in increasing dimensions are formed according to the real roots of those polynomials.  First, the real line is decomposed according to the roots of the polynomials in $S_1$. Then over each cell $c$ in that decomposition, the bivariate polynomials $S_2$ are taken at a sample point and a decomposition of $c \times \R$ is produced according to their roots.  Taking the union gives the decomposition of $\R^2$ and we proceed this way to a decomposition of $\R^n$.  The decompositions are cylindrical (projections of any two cells onto their first $i$ coordinates are either identical or disjoint) and each cell is a semi-algebraic set (described by polynomial relations).  
Collins' original algorithm used a projection operator which guaranteed CADs of $\R^n$ on which the polynomials in $\phi$ had constant sign, and thus $\phi$ constant truth value, on each cell.  Hence only a single sample point from each cell needed to be tested and the equivalent quantifier free formula $\psi$ could be generated from the semi-algebraic sets defining the cells in the CAD of $\R^k$ for which $\Phi$ is true. 

Since the publication of the original algorithm, there have been numerous improvements, optimisations and extensions of CAD (with a summary of the first 20 years given by Collins \cite{Collins1998}). 
Of great importance is the improvement to the projection operator used.  Hong \cite{Hong1990} proved that a refinement of Collins' operator was sufficient and then McCallum \cite{McCallum1998} presented a further refinement which could only be used for input that was \emph{well-oriented} and was in turn improved by Brown \cite{Brown2001a}.  
Further refinements are possible by removing the need for sign-invariance of polynomials while maintaining truth-invariance of a formula, with McCallum \cite{McCallum1999} presenting an operator for use when an equational constraint is present (an equation logically implied by a formula) and Bradford \textit{et al}. \cite{BDEMW13} extending this to the case of multiple formulae. 
Collins and Hong \cite{CH91} described Partial CAD for QE, where lifting over a cell is aborted if there already exists sufficient information to determine the truth of $\phi$ on that cell.  Other recent CAD developments of particular note include the use of symbolic-numeric techniques in the lifting stage \cite{IYAY09, Strzebonski06} and the alternative to projection and lifting offered by decompositions of complex space via regular chains technology \cite{CMXY09}.

When using CAD we have to assign an ordering to the variables (the labels $i$ on the $x_i$ in the discussion above).  This dictates the order in which the variables are eliminated during projection and thus the sub-spaces for which CADs are produced en route to a CAD of $\mathbb{R}^n$. For some applications this order is fixed but for others there may be a free or constrained choice.  When using CAD for QE we must project quantified variables before unquantified ones.  Further, the quantified variables should be projected in the order they occur, unless successive ones have the same quantifier in which case they may be swapped.  The ordering can have a big effect on the output and performance of CAD \cite{BD07, DSS04, BDEW13}.

\subsection{Machine Learning}
\label{SUBSEC:ML}

Machine learning \cite{alpaydin2004introduction} deals with the design of programs that can learn rules from data.  This is often a very attractive alternative to manually constructing them when the underlying functional relationship is very complex. Machine learning techniques have been widely used in many fields, such as web searching\cite{boyan1996machine}, text categorization\cite{sebastiani2002machine}, robotics\cite{stone2000multiagent}, expert systems\cite{forsyth1986machine} and many others.

Various machine learning techniques have been developed. McCulloch and Pitts\cite{mcculloch1943logical} created the first computational model for \emph{neural networks} called \emph{threshold logic}. Following that, Rosenblatt~\cite{rosenblatt1958perceptron} proposed the \emph{perceptron} as an iterative algorithm for supervised classification of an input into one of several possible non-binary outputs. A later development was the \emph{decision tree} \cite{alpaydin2004introduction}, which is a simple representation for classifying examples. The main idea here is to apply serial classifications which refine the output state. At the same time as the \emph{decision tree} was being developed, the \emph{multi-layer perceptron} \cite{hornik1989multilayer} was explored. It is a modification of the standard linear perceptron and can distinguish data that are non-linearly separable. 

In the last decade, the use of machine learning has spread rapidly following the invention of the \emph{Support Vector Machine} (SVM) \cite{scholkopf2004kernel}. This was a development of the perceptron approach and gives a powerful and robust method for both classification and regression. \emph{Classification} refers to the assignment of input examples into a given set of classes (the output being the class labels).  \emph{Regression} refers to a supervised pattern analysis in which the output is real-valued.  The SVM technology can deal efficiently with high-dimensional data, and is flexible in modelling diverse sources of data. The standard SVM classifier takes a set of input data and predicts one of two possible classes from the input. Given a set of examples, each marked as belonging to one of two classes, an SVM training algorithm builds a model that assigns new examples into one of the classes. The examples used to fit the model are called training examples. 

An important concept in the SVM  theory is the use of a kernel function \cite{shawe2004kernel}, which maps data into a high dimensional kernel-defined feature space and then separates samples in the transformed space. Kernel functions enable operations in feature space without ever computing the coordinates of the data in that space. Instead they simply compute the inner products between all pairs of data vectors. This operation is generally computationally cheaper than the explicit computation of the coordinates. 

The machine learning experiment described in this paper uses \textsc{SVM-Light} (see Joachims~\cite{Joachims/99a}) which is an implementation of SVMs in C. The \textsc{SVM-Light} software consists of two programs: \textsc{SVM learn} and \textsc{SVM classify}. \textsc{SVM learn} fits the model parameters based on the training data and user inputs (such as the kernel function and the parameter values). \textsc{SVM classify} uses the generated model to classify new samples.
It calculates a hyperplane of the $n$-dimensional transformed feature space, which is an affine subspace of dimension $n - 1$ dividing the space into two corresponding to the two distinct classes.  \textsc{SVM classify} outputs margin values which are a measure of how far the sample is from this separating hyperplane. Hence the margins are a measure of the confidence in a correct prediction. A large margin represents high confidence in a correct prediction. The accuracy of the generated model is largely dependent on the selection of the kernel functions and parameter values. 

\section{Methodology} 
\label{SEC:Methodology}

\subsection{CAD implementation and heuristics}

For the machine learning experiment we decided to focus on a single CAD implementation, \textsc{Qepcad} \cite{Brown2003b}.  We note that other CAD implementations are available, as discussed further in Section \ref{SEC:Future}.

\textsc{Qepcad} is an interactive command line program written in C for performing \textbf{Q}uantifier \textbf{E}limination with \textbf{P}artial \textbf{CAD}.  It was chosen as it is a competitive implementation of both CAD and QE that also allows the user some control and information during its execution.  We used \textsc{Qepcad} with its default settings which implement McCallum's projection operator \cite{McCallum1998} and partial CAD \cite{CH91}.  It can also makes use of an equational constraint automatically (via the projection operator \cite{McCallum1999}) when one is explicit in the formula, (where \emph{explicit} means the formula is a conjunction of the equational constraint with a sub-formula). 

In the experiment we used three existing heuristics for picking a CAD variable ordering:
\begin{description}
\item[Brown:] This heuristic chooses a variable ordering according to the following criteria, starting with the first and breaking ties with successive ones:
\begin{enumerate}[(1)]
\item Eliminate a variable first if it has lower overall degree in the input.
\item Eliminate a variable first if it has lower (maximum) total degree of those terms in the input in which it occurs.
\item Eliminate a variable first if there is a smaller number of terms in the input which contain the variable.
\end{enumerate}
It is labelled after Brown who suggested it \cite{Brown2004}.
\item[sotd:] This heuristic constructs the full set of projection polynomials for each permitted ordering and selects the ordering whose corresponding set has the lowest sum of total degrees for each of the monomials in each of the polynomials. It is labelled sotd for \emph{sum of total degree} and was suggested by Dolzmann, Seidell and Sturm \cite{DSS04}, whose study found it to be a good heuristic for both CAD and QE by CAD.
\item[ndrr:] This heuristic constructs the full set of projection polynomials for each ordering and selects the ordering whose set has the lowest number of distinct real roots of the univariate polynomials within.  It is labelled ndrr for \emph{number of distinct real roots} and was suggested by Bradford \textit{et al}. \cite{BDEW13}. Ndrr was shown to assist with examples where sotd failed.
\end{description}  
Brown's heuristic has the advantage of being very cheap, since it acts only on the input and checks only simple properties. The ndrr heuristic is the most expensive (requiring real root isolation), but is the only one to explicitly consider the real geometry of the problem (rather than the geometry in complex space).  

All three heuristics may identify more than one variable ordering as a suitable choice.  In this case we took the heuristic's choice to be the first of these after they had been ordered lexicographically.
\footnote{This final choice may depend on the convention used for displaying the variable ordering.  \textsc{Qepcad} and the notes where Brown introduces his heuristic \cite{Brown2004} use the convention of ordering variables from left to right so that the last one is projected first.  On the other hand, \textsc{Maple} and the papers introducing sotd and ndrr \cite{DSS04, BDEW13} use the opposite convention.  The heuristics were implemented in \textsc{Maple} and so ties were broken by picking the first lexicographically on the second convention.  This corresponds to picking the first under a reverse lexicographical order under the \textsc{Qepcad} convention.  The important point is that all three heuristics had ties broken under the same convention and so were treated fairly.}

\subsection{Problem data}

Problems were taken from the nlsat dataset~\cite{nlsat}, chosen over more traditional CAD problem sets (such as Wilson \textit{et al}. \cite{WBD12_EX}) as these did not have sufficient numbers of problems for machine learning.  7001 three-variable CAD problems were extracted for our experiment.  The number of variables was restricted for two reasons. First to make it feasible to test all possible variable orderings and second to avoid the possibility that \textsc{Qepcad} will produce errors or warnings related to well-orientedness with the McCallum projection \cite{McCallum1998}.  

Two experiments were undertaken, applying machine learning to CAD itself and to QE by CAD.  QE is clearly very important throughout engineering and the sciences, but increasingly CAD has been applied outside of this context, as discussed in the introduction.  We performed separate experiments since for quantified problems \textsc{Qepcad} can use the partial CAD techniques to stop the lifting process early if the outcome is already determined, while the full process is completed for unquantified ones and the two outputs can be quite different. 

The problems from the nlsat dataset are all fully existential (satisfiability or SAT problems).  A second set of problems for the quantifier free experiment was obtained by simply removing all quantifiers.  An example of the \textsc{Qepcad} input for a SAT problem is given in Figure \ref{fig:QIn} with the corresponding input for the unquantified problem in Figure \ref{fig:QFIn}.
Of course, for such quantified problems there are better alternatives to building a CAD (see for example the work of Jovanovic and de~Moura \cite{JdM12}).  However, our decision to use only SAT problems was based on availability of data rather than it being a requirement of the technology, and so we focus on CAD only here and discuss how we might generalise our data in Section \ref{SEC:Future}.  
For both experiments, the problems were randomly split into training sets (3545 problems in each), validation sets (1735 problems in each) and test sets (1721 problems in each)
\footnote{The data is available at \texttt{http://www.cl.cam.ac.uk/$\sim$zh242/data}.}.

\begin{figure}[ht]
\setlength{\belowcaptionskip}{5pt plus 3pt minus 2pt}
\caption{Sample \textsc{Qepcad} input for a quantified problem.}
\begin{verbatim}
(x0,x1,x2)
0
(Ex0)(Ex1)(Ex2)[[((x0 x0) + ((x1 x1) + (x2 x2))) = 1]].
go
go
go
d-stat
go
finish
\end{verbatim}
\label{fig:QIn}
\caption{Sample \textsc{Qepcad} input for a quantifier free problem.}
\begin{verbatim}
(x0,x1,x2)
3
[[((x0 x0) + ((x1 x1) + (x2 x2))) = 1]].
go
go
d-proj-factors
d-proj-polynomials
go
d-fpc-stat
go
\end{verbatim}
\label{fig:QFIn}
\end{figure}

\subsection{Evaluating the heuristics}

Since each problem has three-variables and all the quantifiers are the same, all six possible variable orderings are admissible.  For each ordering we had \textsc{Qepcad} build a CAD and measured the number of cells. The best ordering was defined as the one resulting in the smallest cell count, (and if more than one ordering gives the minimal both orderings are considered the best).  The decision to focus on cell counts (rather than say computation time) was made so that our experiment could validate the use of machine learning to CAD theory, rather than just the \textsc{Qepcad} implementation.  Further, it is usually the case that cell counts and timings are strongly correlated.  

The heuristics (Brown, sotd and ndrr) have been implemented in \textsc{Maple} (as part of the freely available \texttt{ProjectionCAD} package \cite{England13b}) and for each problem the orderings suggested by the heuristics were recorded and compared to the cell counts produced by \textsc{Qepcad}  
\footnote{When comparing care must be taken when changing between the different variable ordering conventions (see Footnote 1).}.
Note that all three heuristics do not discriminate on the structure of any quantifiers.  As discussed above, some heuristics are more expensive than others.  However, since none of the costs were prohibitive for our data set they are not considered here.

Machine learning was applied to predict which of the three heuristics will give an \emph{optimal} variable ordering for a given problem, where \emph{optimal} means the lowest cell count of the selected CADs.  Note that in the quantified case \textsc{Qepcad} can collapse stacks when sufficient truth values for the constituent cells have been discovered to determine a truth value for the base cell.  Hence, since our problems are all fully existential, the output for all quantified problems is always a single cell: true or false.  Therefore, in these cases it was not the number of cells in the output that was used but instead the number of cells constructed during the process (hence the statistics commands in Figures \ref{fig:QIn} and \ref{fig:QFIn} differ).

\subsection{Problem features}

To apply machine learning, we need to identify features of the CAD problems that might be relevant to the correct choice of the heuristics. A feature is an aspect or measure of the problem that may be expressed numerically. Table 1 shows the 11 features that we identified, where $(x_{0},x_{1},x_{2})$ are the three variable labels used in all our problems.  
The number of features is quite small, compared to other machine learning experiments.  They were chosen as easily computable features of the problems which could affect the performances of the heuristics.  Other features were considered (such as the maximum coefficient and the proportion of constraints that were equations) but were not found to be useful.
Further investigation into feature selection may be a topic of our future work.  

\begin{table}[h]
  \caption{Description of the features used. The proportion of a variable occurring in polynomials is the number of polynomials containing the variable divided by total number of polynomials. The proportion of a variable occurring in monomials is the number of terms containing the variable divided by total number of terms in polynomials.}
  \label{table:feature}
  \setlength{\tabcolsep}{10pt}
  \def\arraystretch{1.2}%
  \centering
    \begin{tabular}{c l} 
      \hline
      Feature number & Description \\  \hline \hline
      1 & Number of polynomials. \\ 
      2 & Maximum total degree of polynomials. \\ 
      3 & Maximum degree of $x_{0}$ among all polynomials. \\
      4 & Maximum degree of $x_{1}$ among all polynomials. \\
      5 & Maximum degree of $x_{2}$ among all polynomials. \\ 
      6 & Proportion of $x_{0}$ occurring in polynomials. \\
      7 & Proportion of $x_{1}$ occurring in polynomials. \\
      8 & Proportion of $x_{2}$ occurring in polynomials. \\ 
      9 & Proportion of $x_{0}$ occurring in monomials. \\ 
      10 & Proportion of $x_{1}$ occurring in monomials. \\
      11 & Proportion of $x_{2}$ occurring in monomials. \\ \hline
    \end{tabular}
\end{table}

Each feature vector in the training set was associated with a label, $+1$ (positive examples) or $-1$ (negative examples), indicating in which of two classes it was placed.  To take Brown's heuristic as an example, a corresponding training set was derived with each problem labelled $+1$ if Brown's heuristic suggested a variable ordering with the lowest number of cells, or $-1$ otherwise. 

The features could all be easily calculated from the problem input using \textsc{Maple}.  For example. if the input formula is defined using the set of polynomials
\[
\{
-6x_{0}^2-x_{2}^3-1,  \quad
x_{0}^4x_{2}+9x_{1},  \quad
x_{0}+x_{0}^2-x_{2}x_{0}-5
\}
\]
then the problem will have the feature vector 
\[
\left[ 
3, 5, 4, 1, 3, 1, \frac{1}{3}, 1, \frac{5}{9}, \frac{1}{9}, \frac{1}{3} 
\right].
\]
After the feature generation process, the training data (feature vectors) were  normalized so that each feature had zero mean and unit variance across the set.  The same normalization was then also applied to the validation and test sets.

\subsection{Parameter Optimization}
\label{SUBSEC:PO}

\textsc{SVM-Light} was used to do the classification for this experiment. As stated in Section \ref{SUBSEC:ML}, SVMs use kernel functions to map the data into higher dimensional spaces where the data may be more easily separated. \textsc{SVM-Light} has four standard kernel functions: linear, polynomial, sigmoid tanh and radial basis function. For each kernel function, there are associated parameters which must be set. An earlier experiments applying machine learning to an automated theorem prover \cite{UCAM-CL-TR-792} found the radial basis function (RBF) kernel performed well in finding a relation between the simple algebraic features and the best heuristic choice.  Hence the same kernel was selected for this experiment (other kernel functions may be tested in future work).  
The RBF function is defined as:
\[
K(x, x\prime ) = \exp \left( -\gamma ||x - x\prime ||^2 \right)
\]
where $K$ is the kernel function, $x$ and $x\prime$ are feature vectors. There is a single parameter $\gamma$ in the RBF kernel function. Besides the parameter $\gamma$, two other parameters are involved in the SVM fitting process. The parameter $C$ governs the trade-off between margin and training error, and the cost factor $j$ is used to correct imbalance in the training set and we set it equal to the ratio between negative and positive samples. Given a training set, we can easily compute the value of parameter $j$ by looking at the sign of the samples. However, it is not that trivial to find the optimal values of $\gamma$ and $C$. 

In machine learning, \emph{Matthew's correlation coefficient} (MCC) \cite{baldi2000assessing} is often used to evaluate the performance of the binary classifications.  It takes into account true and false positives and negatives:
\[
{\rm MCC} = \frac{{\rm TP}*{\rm TN}-{\rm FP}*{\rm FN}}{\sqrt{({\rm TP}+{\rm FP})({\rm TP}+{\rm FN})({\rm TN}+{\rm FP})({\rm TN}+{\rm FN})}}
\]
In this equation, TP is the number of true positives, TN is the number of true negatives, FP is the number of false positives and FN is the number of false negatives. The denominator is set to $1$ if any sum term is zero. This measure has the value $1$ if perfect prediction is attained, $0$ if the classifier is performing as a random classifier, and $-1$ if the classifier exactly disagrees with the data. 

A grid-search optimisation procedure was used with the training and validation set, involving a search over a range of $(\gamma ,C)$ values to find the pair which would maximize MCC. We tested a commonly used range of value of $\gamma$ (varied between $2^{-15}, 2^{-14}, 2^{-13}, \dots, 2^{3}$) and $C$ (varied between $2^{-5}, 2^{-4}, 2^{-3}, \dots, 2^{15}$) in our grid search process \cite{HSU2003}. Following the completion of the grid-search, the values for kernel function and model parameters giving optimal MCC results were selected for each individual CAD heuristic classifier.  We also performed a similar calculation, selecting parameters to maximise the $F_1$-score \cite{joachims2005support}, but the results using MCC were superior.

The classifiers with optimal $(\gamma ,C)$ were applied to the test set to output the margin values~\cite{cristianini2000introduction}. In an ideal case, only one classifier would return a positive result for any problem, where selecting a best heuristic is just a case of observing which classifier returns a positive result. However, in practice, more than one classifier will return a positive result for some problems, while no classifiers may return a positive for others. Thus, instead we used the relative magnitudes of the classifiers in our experiment. The classifier with most positive (or least negative) margin was selected to indicate the best decision procedure for the selection.  

\section{Results}
\label{SEC:Results}

The experiment was run as described in Section \ref{SEC:Methodology}.  We use the number of problems for which a selected variable ordering is optimal to measure the efficacy of each heuristic separately, and of the heuristic selected by machine learning.   

Table \ref{table:subset} breaks down the results into a set of mutually exclusive outcomes that describe all possibilities.  The column headed `Machine Learning' indicates the heuristic selected by the machine learned model with the next three columns indicating each of the fixed heuristics tested.  For each of these four heuristics, we may ask the question ``Did this heuristic select the optimal variable ordering?''  A `Y' in the table indicates yes and an `N' indicates no, with each of the 13 cases listed covering all possibilities.  Note that at least one of the fixed heuristics must have a `Y' since, by definition, the optimal ordering is obtained by at least one heuristic while if they all have a Y it is not possible for machine learning to fail.  For each of these cases we list the number of problems for which this case occurred for both the quantifier free and quantified experiments.

\begin{table}
  \caption{Categorising the problems into a set of mutually exclusive cases characterised by which heuristics were successful.}
  \label{table:subset}
\centering
  \setlength{\tabcolsep}{6pt}
  \def\arraystretch{1.2}%
    \begin{tabular}{c c c c c c c} 
      \hline 
      Case & Machine Learning & sotd & ndrr & Brown & Quantifier Free & Quantified \\  \hline \hline
      1  & Y  & Y & Y & Y & 399 & 573 \\ \hline
      2  & Y  & Y & Y & N & 146 & 96 \\
      3  & N  & Y & Y & N & 39  & 24  \\ \hline 
      4  & Y  & Y & N & Y & 208  & 232  \\
      5  & N  & Y & N & Y & 35   & 43  \\ \hline 
      6  & Y  & N & Y & Y & 64  & 57  \\
      7  & N  & N & Y & Y & 7  & 11   \\ \hline 
      8  & Y  & Y & N & N & 106  & 66  \\
      9  & N  & Y & N & N & 106  & 75  \\ \hline 
      10 & Y  & N & Y & N & 159 & 101  \\
      11 & N  & N & Y & N & 58  & 89  \\ \hline 
      12 & Y  & N & N & Y & 230 & 208 \\
      13 & N  & N & N & Y & 164 & 146 \\ \hline 
    \end{tabular}
\end{table}

For many problems more than one heuristic selects the optimal variable ordering and the probability of a randomly selected heuristic giving the optimal ordering depends on how many pick it.  For example, a random selection would be successful $1/3$ of the time if one heuristic gives the optimal ordering or $2/3$ of the time if two heuristics do so. 

In Table \ref{table:subset}, case $1$ is where machine learning cannot make any difference as all heuristics are equally optimal.  We compare the remaining cases pairwise. For each pair, the behaviour of the fixed heuristics are identical and the difference is whether or not machine learning picked a winning heuristic (one of the ones with a Y). We see that in each case machine learning succeeds far more often than fails.  For each pair we can compare with a random heuristic selection. For example, consider cases 2 and 3 where sotd and ndrr are successful heuristics and Brown is not.  A random selection would be successful $2/3$ of the time. For the quantifier free examples, machine learned selection is successful $146/(146+39)$ or approximately 79\% of the time, which is significantly better. 

We repeated this calculation for the quantified case and the other pairs, as shown in Table \ref{table:right}.  In each case the values have been compared to the chance of success when picking a random heuristic, and so there are two distinct sets in Table \ref{table:right}: those where only one heuristic was optimal and those where two are.   We see that machine learning did better for some classes of problems than others.  For example in quantifier free examples, when only one heuristic is optimal machine learning does considerably better if that one is ndrr, while if only one is not optimal machine learning does worse if is Brown.  Nevertheless, the machine learning selection is better than random in every case in both experiments.

\begin{table}
  \caption{Proportion of examples where machine learning picks a successful heuristic.}
\centering
  \label{table:right}
  \setlength{\tabcolsep}{10pt}
  \def\arraystretch{1.2}%
    \begin{tabular}{c c c c c} 
      \hline 
      sotd & ndrr & Brown & Quantifier Free & Quantified \\  
      \hline \hline
       Y & Y & N & 79\% (>67\%) & 80\% (>67\%)  \\  
       Y & N & Y & 86\% (>67\%) & 84\% (>67\%)  \\  
       N & Y & Y & 90\% (>67\%) & 84\% (>67\%) \\  \hline
       Y & N & N & 50\% (>33\%) & 47\% (>33\%)  \\  
       N & Y & N & 73\% (>33\%) & 53\% (>33\%)  \\  
       N & N & Y & 58\% (>33\%) & 59\% (>33\%)  \\ \hline 
    \end{tabular}
\end{table}

By summing the numbers in Table \ref{table:subset} in which Y appears in a row for the machine learned selection and each individual heuristic, we get Table \ref{table:qf}. This compares, for both the quantifier free and quantified problem sets, the learned selection with each of the CAD heuristics on their own. 

Of the three heuristics, Brown seems to be the best, albeit by a small margin.  Its performance is a little surprising, both because the Brown heuristic is not so well known (having never been formally published) and because it requires little computation (taking only simple measurements on the input).

\begin{table}
  \caption{Total number of problems for which each heuristic picks the best ordering.}
  \label{table:qf}
  \centering
  \setlength{\tabcolsep}{10pt}
  \def\arraystretch{1.2}%
    \begin{tabular}{lcccc}
      \hline
                            & Machine Learning	& sotd	& ndrr & Brown	\\
      \hline \hline
      Quantifier free       & 1312	            & 1039	& 872  & 1107	\\
      Quantified            & 1333	            & 1109	& 951  & 1270	\\
      \hline
    \end{tabular}
\end{table}

For the quantifier free problems there were 399 problems where every heuristic picked the optimal, 
499 where two did and 
823 where one did.  Hence for this problem set the chances of picking a successful heuristic at random is 
\[
\frac{100}{1721} \left( 399 + 499*\tfrac{2}{3} + 823*\tfrac{1}{3} \right) \simeq 58\%
\]
which compares with $100 * 1312/1721 \simeq 76\%$ for machine learning.
For the quantified problems the figures are 
$64\%$ 
and $77\%$.  
Hence machine learning performs significantly better than a random choice in both cases.  Further, if we were to use only the heuristic that performed the best on this data, the Brown heuristic, then we would pick a successful ordering for approximately 
$64\%$ of the quantifier free problems and 
$74\%$ of the quantified problems.  So we see that a machine learned choice is also superior to using any one heuristic.

\section{Possibilities for extending the experiment}
\label{SEC:Future}

Although a large data set of real world problems was used, we note that in some ways the data was quite uniform.  A key area of future work is experimentation on a wider data set to see if these results, both the benefit of machine learning and the superiority of Brown's heuristic, are verified more generally.  An initial extension would be to relax the parameters used to select problems from the nlsat dataset, for example by allowing problems with more variables.  

One key restriction with this dataset is that all problems have one block of existential quantifiers.  Note that our restriction to this case followed the availability of data rather than any technical limitation of the machine learning.  Possible ways to generalise the data include randomly applying quantifiers to the the existing problems, or randomly generating whole problems.  However, this would mean the problems no longer originate from real applications, and it has been noted in the past that random problems for CAD can be unrepresentative.

We do not suggest SVM as the only suitable machine learning method for this experiment, but overall a SVM with the RBF kernel worked well here. It would be interesting to see if other machine learning methods could offer similar or even better selections.  Further improvements may also come from more work on the feature selection.  The features used here were all derived from the polynomials involved in the input.  One possible extension would be to consider also the type of relations present and how they are connected logically (likely to be particularly beneficial if problems with more variables or more varied quantifiers are allowed).

A key extension for future work will be the testing of other heuristics.  For example the greedy sotd heuristic \cite{DSS04} which chooses an ordering one variable at a time based on the sotd of new projection polynomials or combined heuristics, (where we narrow the selection with one and then breaking the tie with another).  We also note that there are other questions of CAD problem formulation besides variable ordering \cite{BDEW13} for which machine learning might be of benefit.

Finally, we note that there are other CAD implementations.  In addition to \textsc{Qepcad} there is \texttt{ProjectionCAD} \cite{England13b}, \texttt{RegularChains} \cite{CMXY09} and \texttt{SyNRAC} \cite{IYAY09} in \textsc{Maple}, \textsc{Mathematica} \cite{Strzebonski12} and \texttt{Redlog} \cite{DS97a} in \textsc{Reduce}.  Each implementation has its own intricacies and often different underlying theory so it would be interesting to test if machine learning can assist with these as it does with \textsc{Qepcad}.

\section{Conclusions}
\label{SEC:Conclusion}

We have investigated the use of machine learning for making the choice of which heuristic to use when selecting a variable ordering for CAD, and quantifier elimination by CAD.  
The experimental results confirmed our thesis, drawn from personal experience, that no one heuristic is superior for all problems and the correct choice will depend on the problem.  Each of the three heuristics tested had a substantial set of problems for which they were superior to the others and so the problem was a suitable application for machine learning.

Using machine learning to select the best CAD heuristic yielded better results than choosing one heuristic at random, or just using any of the individual heuristics in isolation, indicating there is a relation between the simple algebraic features and the best heuristic choice.  This could lead to the development of a new individual heuristic in the future.

The experiments involved testing heuristics on 1721 CAD problems, certainly the largest such experiment that the authors are aware of.  For comparison, the best known previous study on such heuristics \cite{DSS04} tested with six examples.  We observed that Brown's heuristic is the most competitive for our example set, and this is despite it involving less computation than the others.  This heuristic was presented during an ISSAC tutorial in 2004 (see Brown \cite{Brown2004}), but does not seem to be formally published.  It certainly deserves to be better known.

Finally, we note that CAD is certainly not unique amongst computer algebra algorithms in requiring the user to make such a choice of problem formulation.  More generally, computer algebra systems (CASs) often have a choice of possible algorithms to use when solving a problem.  Since a single formulation or algorithm is rarely the best for the entire problem space, CASs usually use \emph{meta-algorithms} to make such choices, where decisions are based on some numerical parameters \cite{Carette2004}.  These are often not as well documented as the base algorithms, and may be rather primitive.  
To the best of our knowledge, the present paper appears to be the first applying machine learning to problem formulation for computer algebra. The positive results should encourage investigation of similar applications in the field of symbolic computation.

\section*{Acknowledgements}

This work was supported by the EPSRC grant: EP/J003247/1 and the China Scholarship Council (CSC).  The authors thank the anonymous referees for useful comments which improved the paper.

\end{document}